\documentclass[epj]{webofc-mod}
\usepackage[varg]{txfonts}   
\usepackage[hyperindex,breaklinks]{hyperref}
\usepackage{url} 
\def\tev{\textrm{TeV}}
\def\gev{\textrm{GeV}}
\def\pt{\ensuremath{p_{\rm T}}}
\def\met{\ensuremath{E_{\rm T}^{\rm miss}}}

\def\ifb{\ensuremath{\textrm{fb}^{-1}}}
\def\to{\ensuremath{\rightarrow}}
\def\X{\ensuremath{\tilde\chi_1^0}}

\def\t#1{\tilde{ #1}}

\def\mn{\ensuremath{\mu\nu}SSM}
\def\XX{\ensuremath{\tilde\chi_4^0}}
\DeclareGraphicsExtensions{.pdf}
\woctitle{International Conference on New Frontiers in Physics 2013}
\begin{document}
\title{Probing neutrino physics at LHC through $R$-parity breaking \\ supersymmetry}

\author{Vasiliki A.\ Mitsou\inst{1}\fnsep\thanks{\email{vasiliki.mitsou@ific.uv.es}}}

\institute{Instituto de F\'isica Corpuscular (IFIC), CSIC -- Universitat de Val\`encia, \\ 
Parc Cient\'ific de la U.V., C/ Catedr\'atico Jos\'e Beltr\'an 2, \\
E-46980 Paterna (Valencia), Spain}

\abstract{%
$R$-parity conservation is an \emph{ad-hoc} assumption in the most popular versions of supersymmetric scenarios. Hence $R$-parity violation (RPV) not only is allowed but, if induced by bilinear terms (bRPV), it also provides an explanation for the observed neutrino masses and mixing. Here, bRPV models are discussed, giving emphasis on the $\mu$-from-$\nu$ supersymmetric standard model, which is characterised by a rich Higgs sector that easily accommodates a 125-GeV Higgs boson. The phenomenology of such models at the Large Hadron Collider is reviewed and some recent results obtained by LHC experiments are presented. The possibility to extend such analyses to probe bRPV scenarios with a gravitino dark matter candidate is also explored. }
\maketitle
%
\section{Introduction}\label{sc:intro}

Supersymmetry (SUSY)~\cite{susy} is an extension of the Standard Model (SM) which assigns to each SM field a superpartner field with a spin differing by a half unit. SUSY provides elegant solutions to several open issues in the SM, such as the hierarchy problem, the identity of dark matter~\cite{dm-review}, and the grand unification. SUSY is one of the most relevant scenarios of new physics searched at the CERN Large Hadron Collider~\cite{lhc}, yet no signs of SUSY have been observed so far. In view of these null results in \emph{conventional} SUSY searches, it becomes  mandatory to fully explore \emph{non-standard} SUSY scenarios involving $R$-parity violation (RPV)~\cite{rpv} and/or meta-stable particles. 

$R$-parity is defined as: $R = (-1)^{3(B-L)+2S}$, where $B$, $L$ and $S$ are the baryon number, lepton number and spin, respectively. Hence $R=+1$ for all Standard Model particles and $R=-1$ for all SUSY particles. It is stressed that the conservation of $R$~parity is an \emph{ad-hoc} assumption. The only firm restriction comes from the proton lifetime: non-conservation of both $B$ and $L$ leads to rapid proton decay. $R$-parity conservation has serious consequences in SUSY phenomenology in colliders: the SUSY particles are produced in pairs and the lightest supersymmetric particle (LSP) is absolutely stable and weakly interacting, thus providing the characteristic high transverse missing energy (\met) in SUSY events at colliders. Here we highlight two RPV models: bilinear RPV (bRPV) and the $\mu$-from-$\nu$ supersymmetric standard model (\mn), which ---as we shall see--- explain the observed neutrino masses and mixing.

The structure of this paper is as follows. Section~\ref{sc:brpv} is dedicated on bRPV SUSY models: their connection with neutrino physics is explained in Sect.~\ref{sc:brpv-nu}, its phenomenology at the LHC is provided in Sect.~\ref{sc:brpv-pheno}, whilst Sect.~\ref{sc:brpv-atlas} presents the current constraints from the ATLAS experiment. The $\mu$-from-$\nu$ supersymmetric standard model is discussed in Sect.~\ref{sc:munussm}. After a brief review of the theoretical motivation for \mn, descriptions of its possible signatures at the LHC based on multileptons and displaced vertices are given in Sect.~\ref{sc:munussm-multi} and Sect.~\ref{sc:munussm-dv}, respectively. The aspects of RPV SUSY related to dark matter are discussed in Sect.~\ref{sc:dm}. The paper concludes with a summary and an outlook in Sect.~\ref{sc:summary}.

\section{Bilinear $R$-parity violation}\label{sc:brpv}

$R$-parity conservation (RPC) has several consequences such as the stability of the LSP, which is a weakly interacting massive particle (WIMP) and consequently a candidate for dark matter (DM)~\cite{dm-review}. Being a WIMP, once produced at the LHC, it will escape detection, resulting in large missing transverse energy, \met. Providing a DM candidate is one of the strongest advantages of RPC SUSY even though also RPV models exist which may explain DM through, for instance, very light gravitinos~\cite{gravitino,martin,brpv-dm,grefe,steffen,trpv-gravitino}, axions~\cite{axion,steffen} or axinos~\cite{axino,steffen,brpv-axino}. There is no fundamental reason for $R$~parity to be conserved thus lepton and baryon number violating renormalisable terms can appear in the supersymmetric potential in the following way:
\begin{equation}  
W= \lambda_{ijk}   L_i  L_j  E_k^c 
+\lambda_{ijk}'    L_i  Q_j  D_k^c 
+\epsilon_i        L_i  H_u
+\lambda_{ijk}''   U_i^c  D_j^c  D_k^c,
\label{eq:Wsuppot} 
\end{equation} 

\noindent where the couplings $\lambda$, $\lambda'$ and  $\lambda''$  are $3\times 3$ Yukawa matrices ---$i, j, k$ being flavour indexes---, $\epsilon_i$ are parameters with units of mass and $ Q, U, D, L, E, H_u, H_d $ refer to supermultiplets. The first three types of terms lead to lepton number violation, whilst baryon number is violated by the fourth one.

As long as the breaking of $R$~parity is spontaneous, only bilinear terms arise in the effective theory below the RPV scale, thus rendering bilinear $R$-parity violation a theoretically attractive scenario. Moreover, the bilinear model provides a theoretically self-consistent scheme in the sense that trilinear RPV implies, by renormalisation group effects, that also bilinear RPV is present, but not conversely~\cite{Porod:2000hv}. 
 
Therefore the simplest way to break $R$~parity is to add bilinear terms to the MSSM potential. Besides that, additional bilinear soft SUSY breaking terms are introduced, which include small vacuum expectation values for the sneutrinos. The relevant superpotential $W$ and the soft supersymmetry breaking terms $V_{\rm soft}$, which include bilinear $R$-parity violation, would then be~\cite{Romao:1999up}:
\begin{eqnarray}
\qquad\qquad\qquad\qquad\qquad\qquad W &=&  W^{\rm MSSM} + \epsilon_i  \hat{L}_i  \hat{H}_u \\
\qquad\qquad\qquad\qquad\qquad\qquad V_{\rm soft} &=& V_{\rm soft}^{\rm MSSM}- B_i\epsilon_i  \tilde{L}_i  H_u\, ,
\end{eqnarray} 

\noindent where the $B_i$ have units of mass. In fact, if SUSY was not broken, the bilinear terms could be rotated away and be converted into trilinear terms, however the presence of the soft SUSY breaking terms $B_i\epsilon_i  \tilde{L}_i  H_u$ gives bRPV a physical meaning~\cite{Romao:1999up}. 

There are nine new parameters introduced in this model:  $\epsilon_i$, $B_i$ and $v_i$, the latter being the sneutrino vacuum expectation values (vev's). However, after electroweak symmetry breaking and taking into account constraints from neutrino oscillation data, only one free parameter remains in the model, which is set to be of the same order as the others. 

Besides LHC, a possible signal of bRPV SUSY can be observed at a future $e^+e^-$ collider. Using realistic simulated data for the beams and the detector, it has been shown that a discovery of $5\sigma$ significance is feasible, accompanied by a precise ${\mathcal O}(0.1\%)$ measurement of the neutralino LSP mass with an integrated luminosity of 500~\ifb\ at $\sqrt{s}=500~\gev$~\cite{brpv-ilc}. 

\subsection{bRPV and neutrino physics}\label{sc:brpv-nu}

Sneutrino vev's introduce a mixing between neutrinos and neutralinos, leading to a see-saw mechanism that gives mass to one neutrino mass scale at tree level. The second neutrino mass scale is induced by loop effects~\cite{Hirsch:2000ef,Diaz:2003as}. The same vev's are also involved in the decay of the LSP, which in this case is the lightest neutralino. This implies a relation between neutrino physics and some features of the LSP modes. In particular, they are related through the following relation~\cite{Hirsch:2000ef}:
\begin{equation}
\label{tetatm}
\tan^2\theta_{\rm atm} = \left|{\frac{\Lambda_{\mu}}{\Lambda_{\tau}}}\right|^2 
\simeq \frac{BR(\tilde\chi_1^0\rightarrow\mu^\pm W^\mp)}{BR(\tilde\chi_1^0\rightarrow\tau^\pm W^\mp)} 
= \frac{BR(\tilde\chi_1^0\rightarrow\mu^\pm q \bar{q}')}{BR(\tilde\chi_1^0\rightarrow\tau^\pm q \bar q')},
\end{equation}
 
\noindent where $\theta_{\rm atm}$ is the atmospheric neutrino mixing angle and the alignment parameters $\Lambda_i$ are defined as $\Lambda_i = \mu v_i + v_d \epsilon_i$, with $v_d$ the vev of $H_0^d$. This relation between RPV and neutrino physics allows setting bounds on bRPV parameters from results of neutrino experiment and astrophysical data~\cite{Abada:2000xr}. In the opposite direction, a possible positive signal observed in colliders may lead to the determination of some of the bRPV phenomenological properties, which in turn can constrain neutrino-physics parameters~\cite{brpv-nu}.

Equation~(\ref{tetatm}) illustrates the relation between neutrino physics parameters and quantities measurable at colliders such as LHC, emphasising the motivation behind the selection of the specific model. Indeed it has been demonstrated that the ratio on the r.h.s.\ of Eq.~(\ref{tetatm}) can be determined with a precision of 4\% at a 500~\gev\ ILC~\cite{brpv-ilc}. Nevertheless, similar decays of the neutralino to a muon or a tau and two jets are predicted, e.g., when the trilinear couplings $\lambda'_{2jk}$ and $\lambda'_{3jk}$, respectively, are considered non zero.

\subsection{bRPV phenomenology at the LHC}\label{sc:brpv-pheno}

In the particular bilinear $R$-parity violating model considered here, the LSP is the lightest neutralino, $\tilde\chi_1^0$~\cite{deCampos:2007bn}. The bRPV terms are embedded in the minimal Supergravity (mSUGRA) model, which imposes some restrictions which reduce the large number of parameters of the MSSM. The number of parameters in mSUGRA is reduced to five, namely $m_0$, the scalar mass; $m_{1/2}$, the gaugino mass; $A_0$, the trilinear scalar coupling; $\tan \beta = v_u / v_d$, the ratio of the Higgs vev's; and ${\rm sgn}\,\mu $, the sign of the higgsino mass parameter.

The six bRPV parameters for each mSUGRA point are determined by the \texttt{SPheno}~\cite{spheno} spectrum calculator. \texttt{SPheno} uses as input the mSUGRA parameters and the neutrino physics constraints, and delivers as output the bRPV parameters (together with the mass spectra and the decay modes) compatible with these constraints. Once those quantities are calculated for a given set of bRPV parameters within an mSUGRA benchmark point, they can subsequently passed to an event generator and produce collision events at the LHC. 

It is stressed that the sparticle spectrum for bRPV-mSUGRA is ---within theoretical uncertainties--- the same as in RPC mSUGRA; its the LSP decay modes and its lifetime that depend on the bRPV parameters. The \X\ decay length and typical decay modes are shown in Fig.~\ref{fg:brpv-decay}. The neutralino decays are dominated by leptonic $(e,\,\mu)$ or $\tau$ channels, making lepton-based searches ideal for constraining this model. The decay to a muon and a hadronically-decaying $W$ has been studied in detail and it has been shown that it may lead to the measurement of the \X\ mass through the reconstruction of the $\X\to\mu q'\bar{q}$ peak in the $\mu$jj invariant mass~\cite{int-note,emma,brpv-nu}. Furthermore the fact that in the low-$m_{1/2}$ high-$m_0$ region the \X\ is slightly long lived opens the possibility to use searches for displaced vertices~\cite{atlas-dv} in order to probe this model at the LHC~\cite{brpv-dv,brpv-gmsb} or the Tevatron~\cite{brpv-dv-tev}. Lastly, the \X\ decays to one or two neutrinos give rise to a moderately high \met, thus rendering some of the standard \met-based analyses pertinent to bRPV, as we shall see in the next section.
 
\begin{figure}[htb]
\centering
\includegraphics[width=0.5\textwidth]{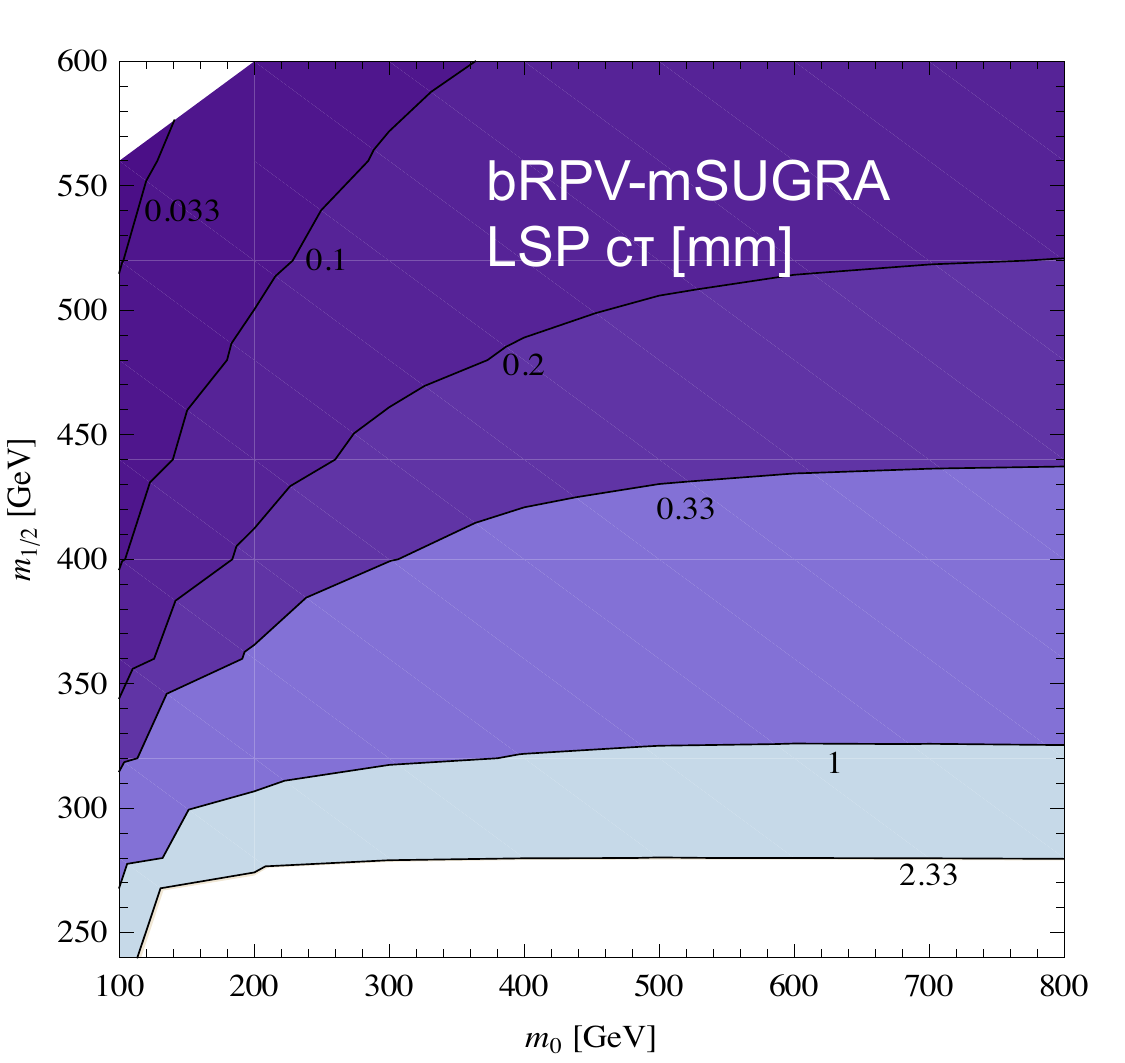}\hfill
\includegraphics[width=0.45\textwidth]{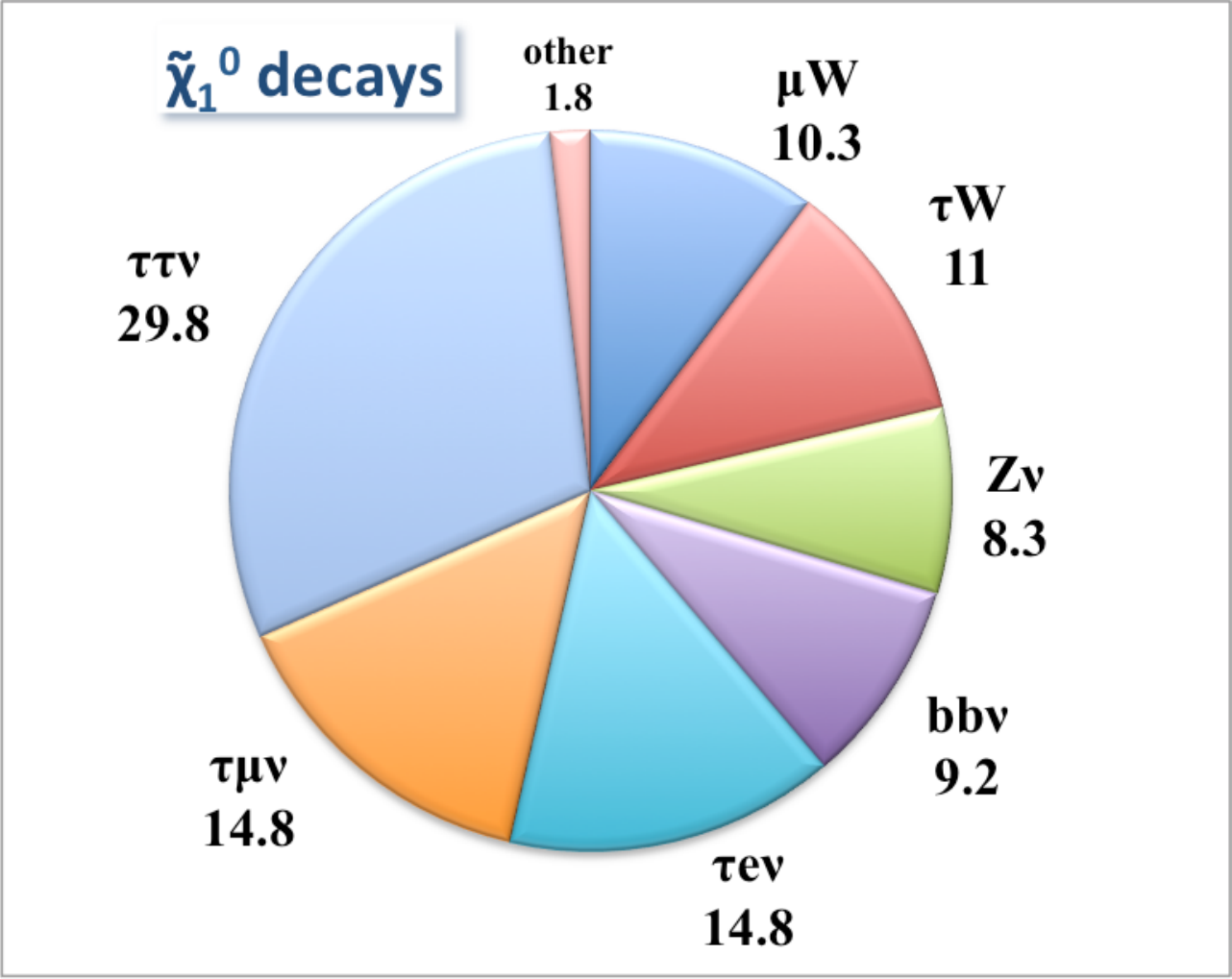}
\caption{{\it Left:} Proper decay length $c\tau$ in millimetres of the lightest neutralino in bRPV-mSUGRA parameter plane $(m_0,\,m_{1/2})$. {\it Right:} Typical \X\ decay modes for a selected bRPV-mSUGRA point.}
\label{fg:brpv-decay}      
\end{figure}  

So far we have discussed the bRPV in the context of supergravity models. If the bRPV terms are introduced to anomaly-mediated SUSY breaking (AMSB) instead, the \X\ LSP decay modes do not change, whereas the wino-like neutralino (as opposed to the bino-like in mSUGRA) is characterised by stronger interactions, hence it is easier to be produced at LHC. The mass degeneracy of the \X\ and the $\t{\chi}^{\rm}_1$ makes the $\t{\chi}^{\rm}_1$ long lived, decaying dominantly through RPV couplings to $\ell\ell\ell$, $\tau\ell\ell$, $\ell b\bar{b}$, $\tau b\bar{b}$~\cite{brpv-amsb}. In the case of gauge-mediated SUSY breaking (GMSB) the \X\ NLSP features delayed ---due to the weak coupling to the LSP--- decays to a graviton and a $Z$, $\gamma$ or $h^0$~\cite{brpv-gmsb}. In other words, the common feature of bRPV phenomenology is the presence of delayed decays leading to interesting signatures at the LHC: displaced vertices, non-pointing photons, etc.

\subsection{Constraints from ATLAS}\label{sc:brpv-atlas}

As mentioned above, thanks to the abundant neutrino production from the LSP decay, bRPV events at the LHC are expected to feature relatively (when compared to Standard Model processes) high missing transverse energy. High lepton/$\tau$ multiplicity is also expected from the LSP decays and from upstream lepton production in the SUSY cascade decay. Both features are exploited when looking for a signal of the bRPV-mSUGRA model, as demonstrated in Ref.~\cite{int-note}, where a first detailed Monte-Carlo-based study on the observability of the $\X\to\mu q'\bar{q}$ at the LHC has been carried out. The very first bounds set on a bilinear RPV model in colliders were provided by an inclusive search for high \met, three or four jets and one muon at $\sqrt{s}=7~\tev$ and $\sim 1~\ifb$ of ATLAS data~\cite{brpv-atlas1,emma}. The \X\ decay to muonic channels is enhanced when compared to electrons, as shown in Fig.~\ref{fg:brpv-decay}, hence muon channels yield stronger limits than their electron counterparts. 

The latest exclusion limits set on this model, when the bilinear terms are embedded in mSUGRA, were obtained with 5~\ifb\ of ATLAS data at $\sqrt{s}=7~\tev$ by searching for events with high jet multiplicity (at least seven jets), large \met\ and exactly one lepton~\cite{brpv-atlas2} and are shown in Fig.~\ref{fg:brpv-atlas}. There is a slight loss of sensitivity in the high-$m_0$ low-$m_{1/2}$ region due to the large LSP lifetime, which causes signal muons to be rejected by the cosmic-muon veto, i.e.\ the cut on muons with high impact parameter. The insignificant ($\sim 2\sigma$) excess of observed events accounts for the weaker-than-expected observed limits, which improve however considerably the previous ones. Ongoing work continues on other channels: $Z$ + jets + \met;  2~same-sign leptons + \met; 0-lepton + jets + \met, $\tau$ + \met, where a mass of the lightest Higgs equal to 125~\gev\ is assumed.

\begin{figure}[htb]
\centering
\sidecaption
\includegraphics[width=0.51\textwidth]{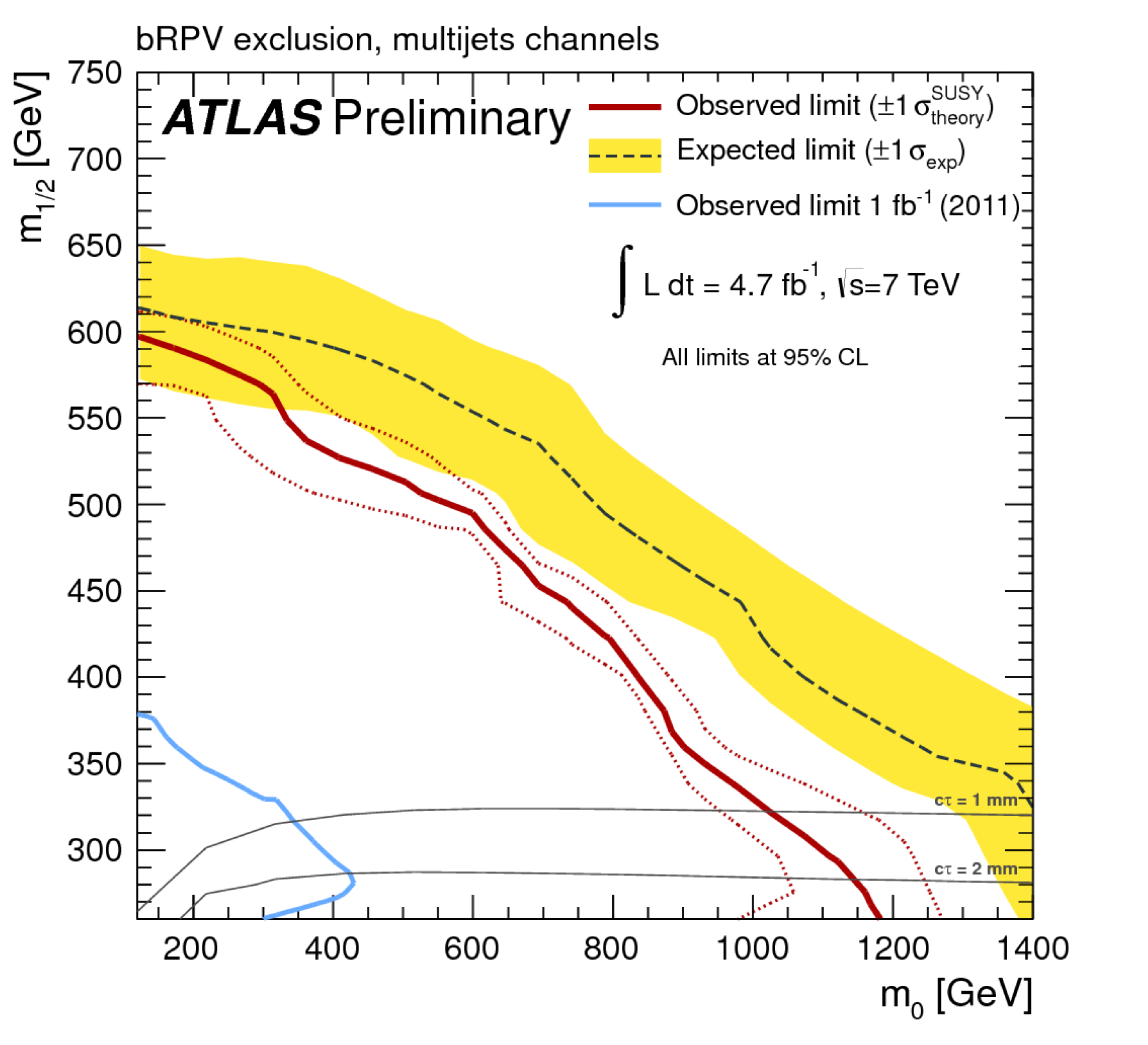}
\caption{Expected and observed 95\% CL exclusion limits in the bilinear $R$-parity violating model obtained by the one-lepton plus multi-jets plus \met\ analysis. The results are obtained by combining the electron and muon channels. The band around the median expected limit shows the $\pm 1\sigma$ variations on the median expected limit, including all uncertainties except theoretical uncertainties on the signal. The dotted lines around the observed limit indicate the sensitivity to $\pm 1\sigma$  variations on these theoretical uncertainties. The thin solid black contours show the LSP lifetime. The result from the previous ATLAS search~\cite{brpv-atlas1,emma} for this model is also shown. From Ref.~\cite{brpv-atlas2}.} \label{fg:brpv-atlas}
\end{figure}

\section{$\mu\nu$SSM: $\mu$-from-$\nu$ supersymmetric standard model}\label{sc:munussm}

The \mn~\cite{mnssm-proposal,mnssm-spectrum} is a supersymmetric standard model that solves the $\mu$~problem~\cite{mu-problem} of the minimal supersymmetric standard model (MSSM) using the $R$-parity breaking couplings between the right-handed neutrino superfields and the Higgs bosons in the superpotential, $\lambda_i\hat{\nu}_i^c\hat{H}_d\hat{H}_u$. The $\mu$~term is generated spontaneously through sneutrino vacuum expectation values, $\mu = \lambda_i\langle\t{\nu}_i^c\rangle$, once the electroweak symmetry is broken, without introducing an extra singlet superfield as in the case of the next-to-MSSM (NMSSM)~\cite{nmssm}. 

In addition, the couplings $\kappa_{ijk}\hat{\nu}_i^c\hat{\nu}_j^c\hat{\nu}_k^c$ forbid a global $U(1)$ symmetry avoiding the existence of a Goldstone boson, and also contribute to spontaneously generate Majorana masses for neutrinos at the electroweak scale. The latter feature is unlike the bilinear RPV model, where only one mass is generated at the tree level and loop corrections are necessary to generate at least a second mass and a neutrino mixing matrix compatible with experiments. In the bilinear model, the $\mu$-like problem~\cite{mu-problem-nilles} is also augmented with three bilinear terms.

In the $\mu\nu$SSM, as a consequence of $R$-parity violation, all the neutral fermions (scalars) mix together and there are ten neutralinos and five charginos (eight $CP$-even and seven $CP$-odd neutral Higgs bosons) mass eigenstates. Analyses of the $\mu\nu$SSM, with attention to the neutrino and LHC phenomenology have also been addressed in Refs.~\cite{mnssm-spectrum,mnssm-others}. Other analyses concerning cosmology such as gravitino dark matter and electroweak baryogenesis can be found in Refs.~\cite{mnssm-gravitino1,mnssm-gravitino2} and in Ref.~\cite{mnssm-baryo}, respectively.


Thus the $\mu\nu$SSM is a well motivated SUSY model with enriched phenomenology and notable signatures, which definitely deserve rigorous analyses by the LHC experiments. Its enlarged Higgs sector can easily accommodate a 125~\gev\ Higgs boson in accordance with the recent discovery of a new scalar boson of such mass by the ATLAS~\cite{ATLAShiggs} and CMS~\cite{CMShiggs} collaborations. 

Here a dedicated collider analysis together with detector simulation of an intriguing signal in the \mn\ featuring non-prompt multileptons at the LHC, arising from the beyond SM decay of a 125~\gev\ scalar into a pair of lightest neutralinos, \XX, is presented~\cite{my-mnssm}. Since $R$~parity is broken, each \XX\ decays into a scalar/pseudoscalar $(h/P)$ and a neutrino, with the $h/P$ further driven to decay into $\tau^+ \tau^-$, giving rise to a $4\tau$ final state, as shown in Fig.~\ref{fg:decay-chain}. The small $R$-parity breaking coupling of \XX\ renders it long~lived, yet it decays inside the inner tracker, thereby yielding clean detectable signatures: (i) high lepton multiplicity; and (ii) charged tracks originating from displaced vertices (DVs). 

\begin{figure}[htb]
\centering
\sidecaption
\includegraphics[width=0.4\textwidth]{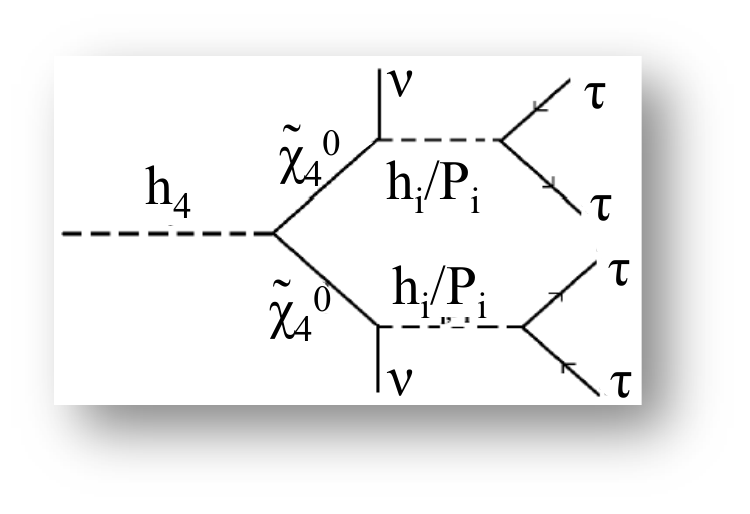}
\caption{The \mn\ decay chain studied in Ref.~\cite{my-mnssm}. The Higgs boson is produced through gluon fusion and has a mass of 125~\gev. The neutralino \XX\ is long-lived and gives rise to displaced $\tau$~leptons.} \label{fg:decay-chain}
\end{figure}

\subsection{Signatures at LHC: multileptons}\label{sc:munussm-multi}

The $\mu\nu$SSM is characterised by the production of several high-\pt\ leptons~\cite{my-mnssm}. Electrons and muons are produced through the leptonic $\tau$ decays, yet muon pairs can appear directly through $h_i/P_i$ decays as well. With the chosen decay mode, the $\tau$ multiplicity is considerable even though the $\tau$-identification efficiency is much lower ($\sim50\%$) when compared to that of electrons and muons ($\gtrsim95\%$). Occasionally highly collimated QCD jets can fake hadronic $\tau$~leptons, $\tau_\text{had}$, and, as a result, $\tau$ multiplicity may exceed the expected number of four. This faking, however disappears with a higher \pt~cut, which should also be sufficient to provide a single-lepton trigger for the analysis. 

Such analyses, apart from the requirement of at least three or four leptons (including taus), require a high value of \met\ and/or of the scalar sum of reconstructed objects: leptons, jets and/or \met. For the chosen signal many neutrinos ($\geq 6$) appearing in the final state from \XX\ and from $\tau$ decay give rise to moderately high ---when compared to signals from RPC supersymmetry--- missing transverse energy, \met, as depicted in Fig.~\ref{fg:met} (left). Besides \met, the scalar sum of the \pt\ of all reconstructed leptons, $H{\rm _T^\ell}$, is also large in such events, as shown in Fig.~\ref{fg:met} (right). Alternatively, the sum of \met\ and $H{\rm _T^\ell}$ can be deployed for further background rejection. These observables can provide additional handles when selecting events with many leptons. In addition, the invariant masses, $m_{\ell^+\ell^-}$ and $m_{2\ell^+2\ell^-}$ may prove useful for the purpose of signal distinction.

\begin{figure}[htbp]
\centering 
\includegraphics[width=0.46\textwidth]{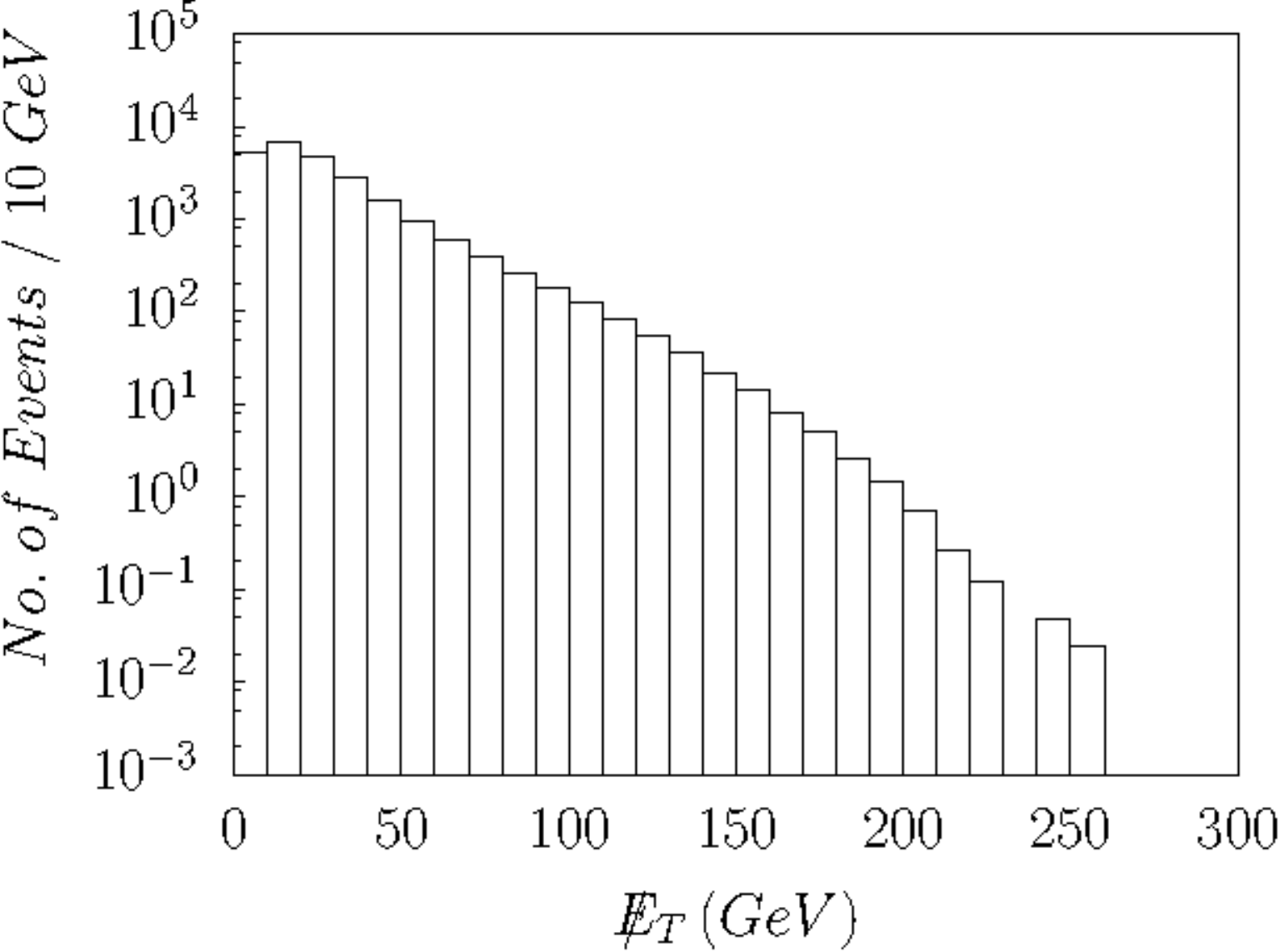}\hfill
\includegraphics[width=0.54\textwidth]{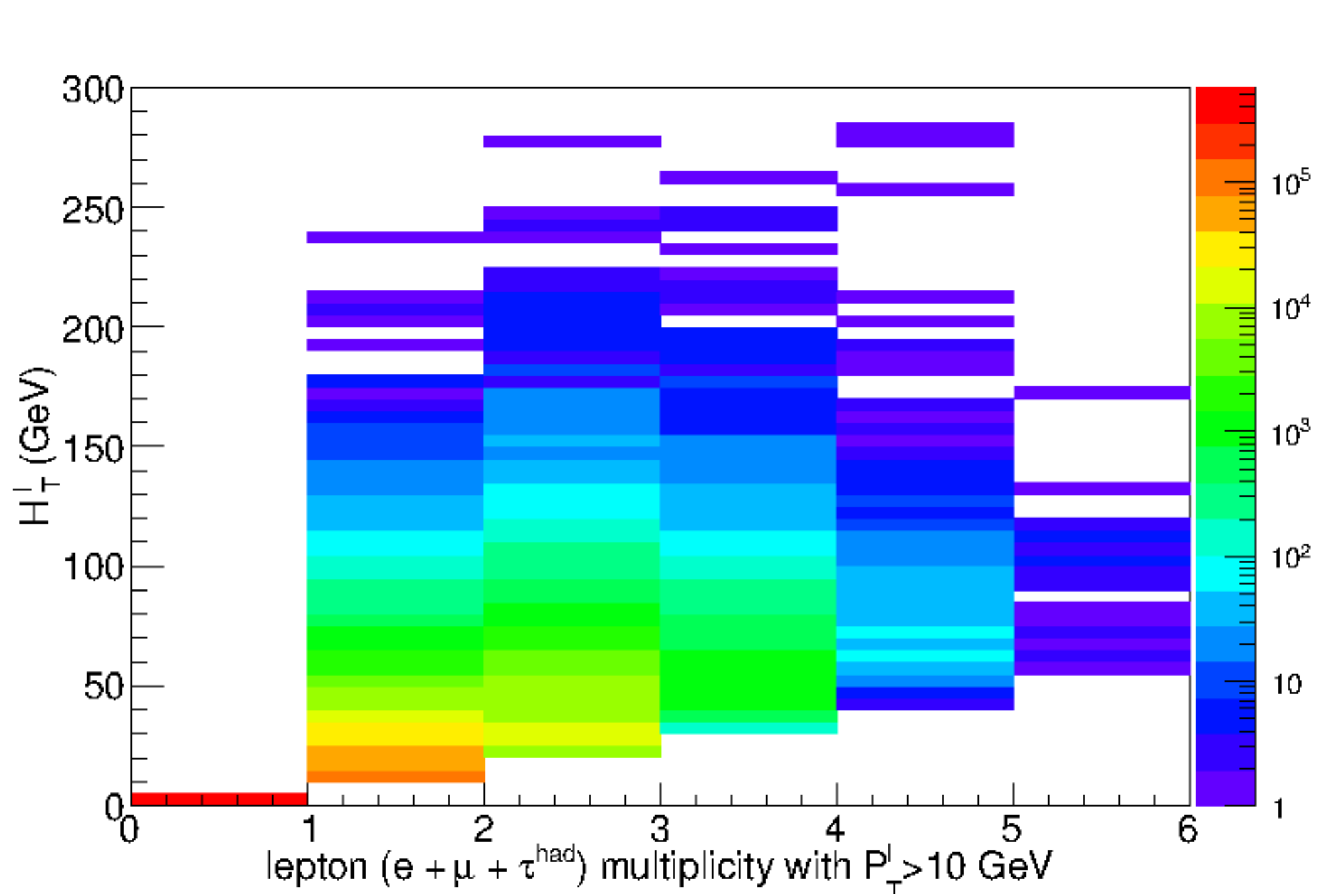}
\caption{\met~distribution (left) and $H{\rm _T^\ell}$ versus lepton multiplicity (right) for an LHC energy of $\sqrt{s}=8~\tev$ and an integrated luminosity $\mathcal{L}=20~\ifb$ for a selected \mn\ point~\cite{my-mnssm}.}
\label{fg:met}
\end{figure}

\subsection{Signatures at LHC: displaced vertices}\label{sc:munussm-dv}

In the benchmark scenario under study, the \XX\ is characterised by a proper lifetime $\tau_{\XX}\approx10^{-9}$~s, which corresponds to a proper decay length of $c\tau_{\XX}\approx 30$~cm. This feature ---quantified in the left panel of Fig.~\ref{fg:dv}, were the decay-length distribution is plotted--- gives rise to displaced vertices. In a significant percentage of events, the \XX\ decays inside the inner tracker of the LHC experiments, e.g.\ 28\% of events decay within $30~{\rm cm}$ and $44$\% events within 1~m. Therefore, the $\mu\nu$SSM signal events will be characterised by displaced $\tau$~leptons plus neutrinos. This distinctive signature opens up the possibility to exploit current or future variations of analyses carried out by ATLAS and CMS looking for a displaced muon and tracks~\cite{atlas-dv} or searching for displaced dileptons~\cite{cms-dileptons}, dijets~\cite{cms-dijets} or  muon jets~\cite{atlas-dl} arising in Higgs decays to pairs of long-lived invisible particles. 

\begin{figure}[htbp]
\centering
\includegraphics[width=0.39\textwidth]{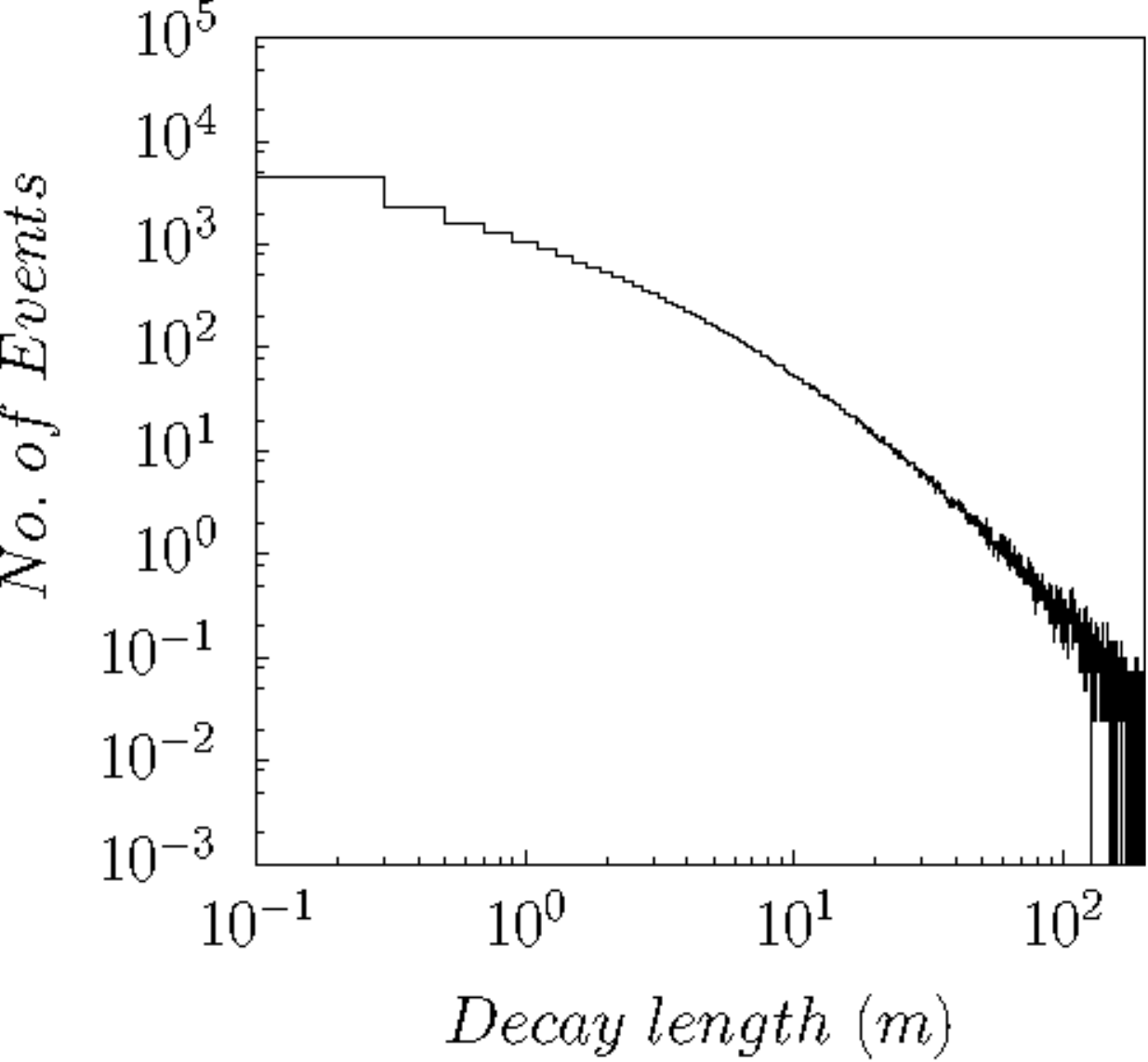}\hfill
\includegraphics[width=0.57\textwidth]{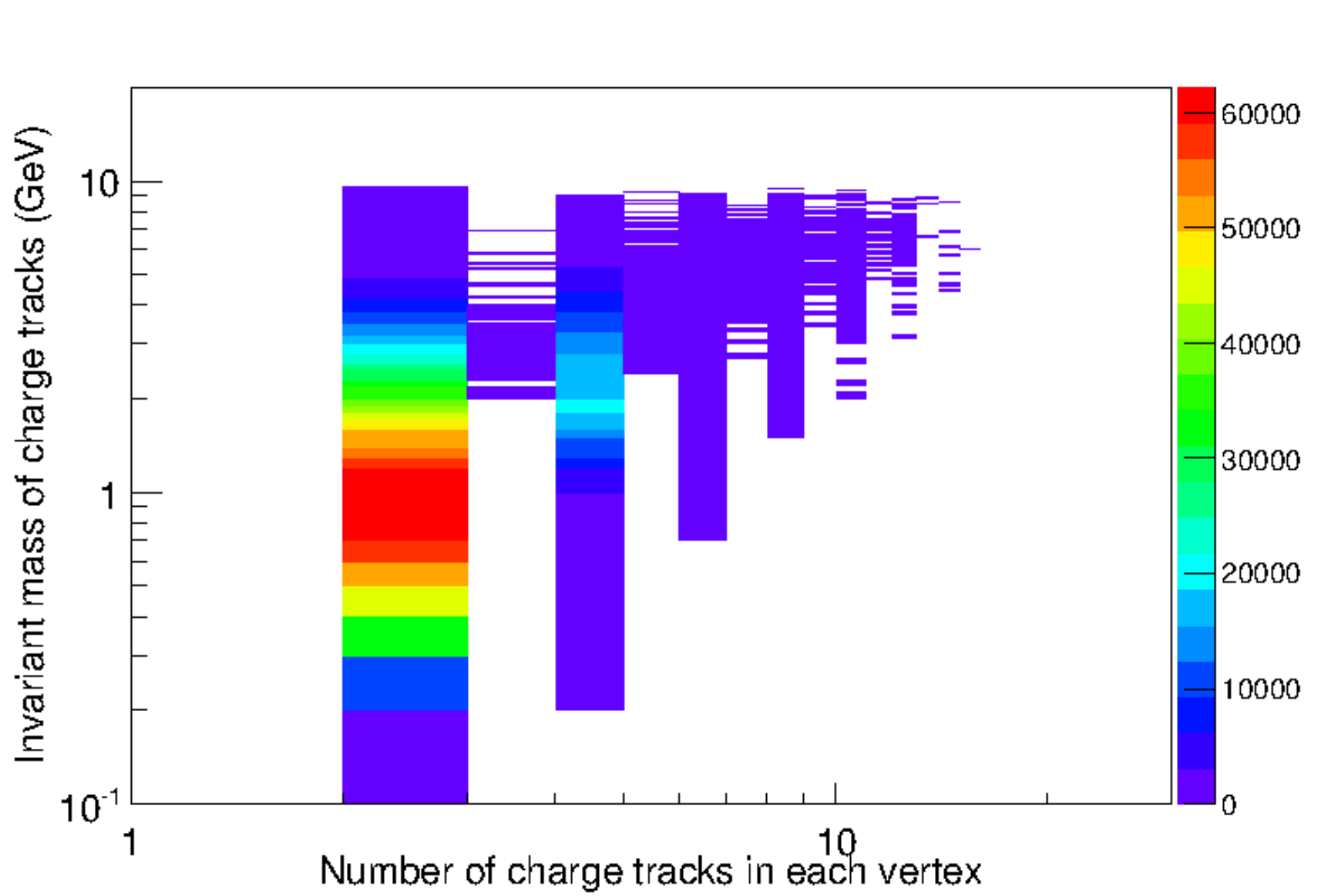}
\caption{\XX\ decay-length distribution (left) and charged-track mass versus the number of charge particles in each vertex (right) for LHC energy of $\sqrt{s}=8~\tev$ and integrated luminosity $\mathcal{L}=20~\ifb$ for a selected \mn\ point~\cite{my-mnssm}.}
\label{fg:dv}
\end{figure}

The kinematics of the DVs and their products in the chosen \mn\ benchmark point have been studied thoroughly in Ref.~\cite{my-mnssm}. The spacial distribution of DVs shows that an appreciable fraction of them falls in the inner-tracker volume of the LHC experiments, i.e.\ $\rho_{\text{DV}}\lesssim 1$~m and $|z_{\text{DV}}|\lesssim 2.5$~m, thus DVs arising in the $\mu\nu$SSM should be detectable at LHC, in principle, either with existing analyses~\cite{atlas-dv,cms-dileptons,cms-dijets,atlas-dl} or via variations of those looking for displaced taus and \met. The average \XX\ boost, on the other hand, as expressed by $\beta\gamma$, where $\beta$ is \XX\ velocity over $c$ and $\gamma$ the Lorentz factor, is comparable to the signal analysed in an ATLAS related search for a muon and tracks originating from DVs~\cite{atlas-dv}. The boost affects the efficiency with which such a DV can be reconstructed, since high \XX\ boost leads to collimated tracks difficult to differentiate from primary vertices. In the right-hand side of Fig.~\ref{fg:dv}, the correlation between the number of charged tracks in each DV, $N_{\rm trk}$, and their invariant mass, $m_{\rm DV}$, is shown. It has been demonstrated~\cite{atlas-dv} that a selection of high-$N_{\rm trk}$ and high-$m_{\rm DV}$ efficiently suppresses background from long-lived SM particles ($B$~mesons, kaons). The modulation observed in $N_{\rm trk}$ is due to the one-prong or three-prong hadronic $\tau$ decays. 

\section{Dark matter and $R$-parity violating SUSY}\label{sc:dm}

We address here the issue of (not necessarily cold) dark matter in SUSY models with $R$-parity violation. It has been shown that these seemingly incompatible concepts \emph{can} be reconciled in bRPV models with a gravitino~\cite{gravitino,martin,brpv-dm} or an axino~\cite{axino} LSP with a lifetime exceeding the age of the Universe. In both cases, RPV is induced by bilinear terms in the superpotential that can also explain current data on neutrino masses and mixings without invoking any GUT-scale physics. Decays of the next-to-lightest superparticle occur rapidly via RPV interactions, and thus they do not upset the Big-Bang nucleosynthesis, unlike the $R$-parity conserving case~\cite{leptog}. Decays of the NLSP into the gravitino and Standard Model particles do not contribute to the gravitino relic density in scenarios with broken $R$~parity. This is because decay processes involving a gravitino in the final state are suppressed compared to $R$-parity breaking decays unless the amount of $R$-parity violation is extremely small.

Interestingly, gravitino decays may produce monochromatic photons via $\t{G}\to\gamma\nu$, opening therefore the possibility to test this scenario with astrophysical searches. Requiring the model parameters to correctly account for observed neutrino oscillation parameters implies that expected rates for $\gamma$-ray lines produced by gravitino decays of mass above a few GeV would be confronted with the Fermi-LAT satellite observations, leading to an upper bound on the gravitino DM mass. For instance, the bRPV parameter region allowed by $\gamma$-ray line searches, dark matter relic abundance, and neutrino oscillation data, has been determined obtaining a limit on the gravitino mass $m_{\t{G}}\lesssim 1-10~\gev$ corresponding to a relatively low reheat temperature~\cite{brpv-dm}. The allowed region in the gravitino mass versus gravitino lifetime plane is shown in the left-hand side of Fig.~\ref{fg:dm}. The yellow region is excluded by $\gamma$-ray line searches: Fermi and EGRET constraints are above and below 1~\gev, respectively.

\begin{figure}[htbp]
\centering
\includegraphics[width=0.49\textwidth]{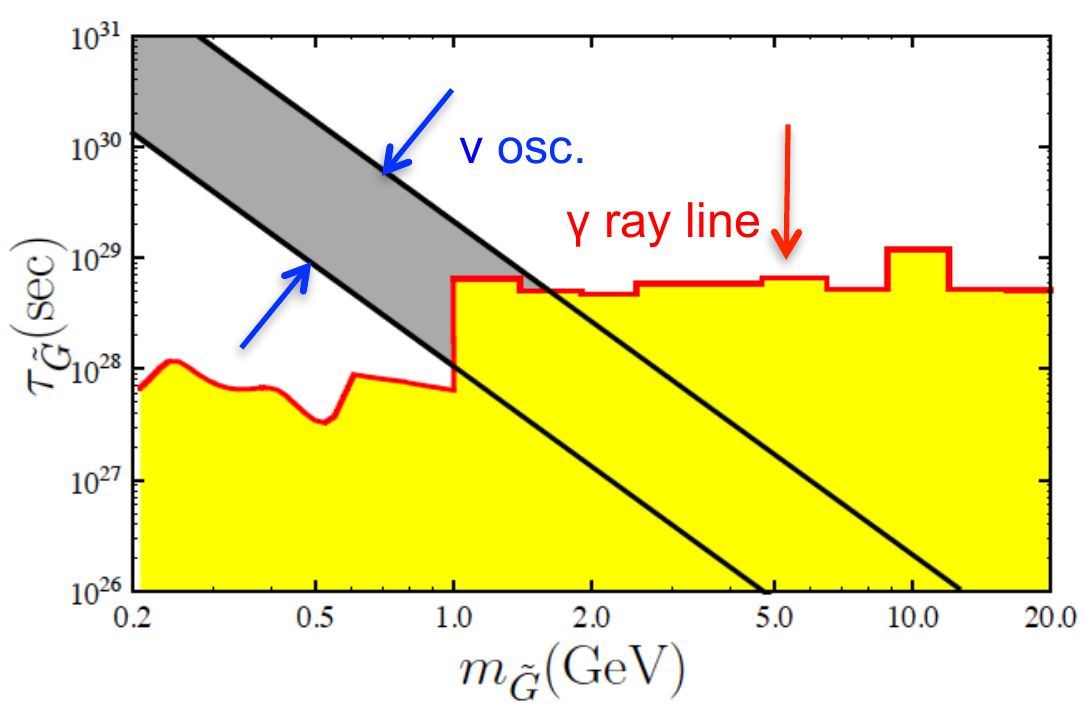}\hfill
\includegraphics[width=0.49\textwidth]{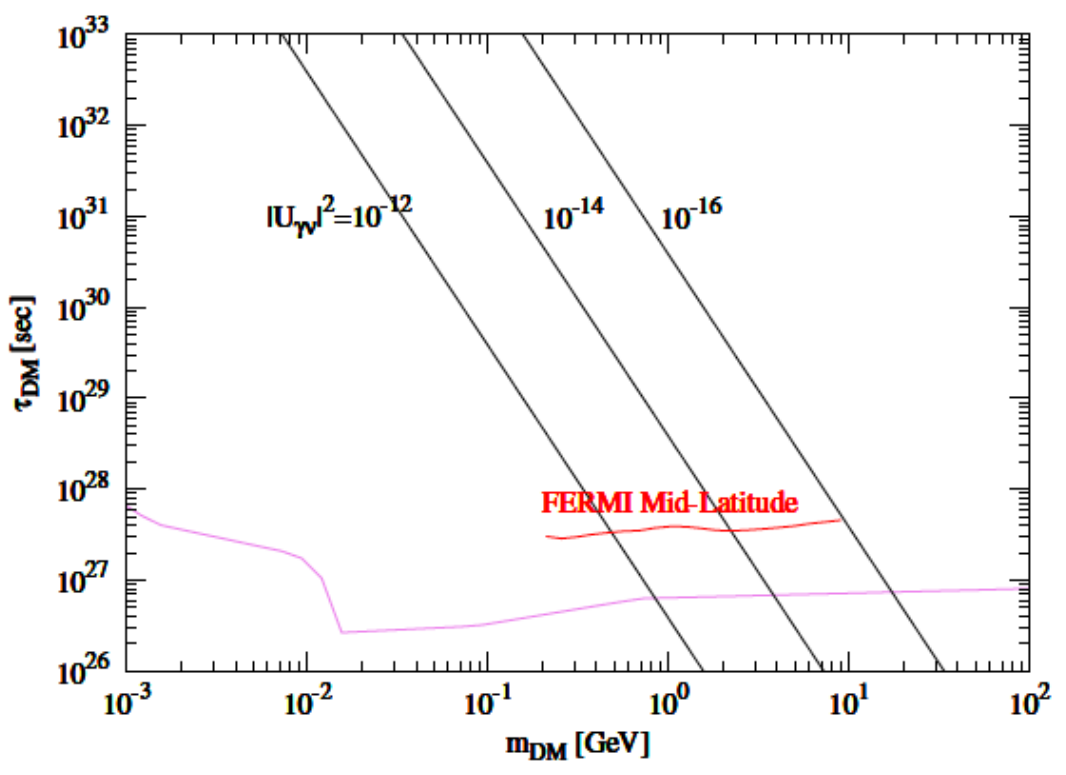}
\caption{{\it Left:} Allowed gravitino mass-versus-lifetime region (grey color) consistent with neutrino oscillation data and astrophysical bounds on $\gamma$-ray lines from dark matter decay for a bilinear RPV model. The lower and upper black lines correspond to $m_{1/2} = 240$ and $3000~\gev$, respectively. Adapted from Ref.~\cite{brpv-dm}. {\it Right:} Constraints on lifetime versus mass for gravitino dark matter in the \mn. The region below the magenta solid line is excluded by several gamma-ray observations. The region below the red solid line is disfavoured by Fermi. Black solid lines correspond to the predictions of the \mn\ for several representatives values of $|U_{\t{\gamma}\nu}|^2 = 10^{-16}-10^{-12}$~\cite{mnssm-gravitino1}. }
\label{fg:dm}
\end{figure}

Recent evidence on the four-year Fermi data that have found excess of a 130~\gev\ gamma-ray line from the Galactic Center (GC)~\cite{fermi130} have been studied in the framework of $R$-parity breaking SUSY. A decaying axino DM scenario based on the SUSY KSVZ axion model with the bilinear $R$-parity violation explains the Fermi 130~\gev\ gamma-ray line excess from the GC while satisfying other cosmological constraints~\cite{brpv-axino}. On the other hand, gravitino dark matter with trilinear RPV ---in particular models with the $LLE$ RPV coupling--- can account for the gamma-ray line, since there is no overproduction of anti-proton flux, while being consistent with big-bang nucleosynthesis and thermal leptogenesis~\cite{trpv-gravitino}.

Measurements of the cosmic-ray antiproton flux by the PAMELA experiment have been used recently to constrain the (decaying) gravitino mass and lifetime in the channels $Z\nu$, $W\ell$ and $h\nu$~\cite{grefe}. Subsequently upper limits have been set on the size of the $R$-parity violating coupling in the range of $10^8$ to $8\times10^{13}$. These limits turn out to be more stringent than those coming from contributions to  neutrino masses or from the baryon asymmetry in the early Universe. Combining them with lower limits from big-bang nucleosynthesis constraints on the NLSP lifetime allows narrowing down the available parameter space for gravitino DM with bRPV.

Such gravitino DM is also proposed in the context of \mn\ with profound prospects for detecting $\gamma$ rays from their decay~\cite{mnssm-gravitino1}. The constraints on $\t{G}$ lifetime versus mass set by Fermi and $\gamma$-ray observations are summarised in the right panel of Fig.~\ref{fg:dm}. If the RPV parameter $|U_{\t{\gamma}\nu}|$ from the $7\times7$ neutralino mixing matrix is between $10^{-16}$ and $10^{-12}$, the correct neutrino masses are reproduced.  Values of the gravitino mass larger than 10~\gev\ are disfavoured, as well as lifetimes smaller than ${\mathcal O}(10^{27}~{\rm s})$.

Further studies on the prospects of the Fermi-LAT telescope to detect such monochromatic lines in five years of observations of the most massive nearby extragalactic objects have been performed~\cite{mnssm-gravitino2}.  It was found that a gravitino with a mass range of $0.6-2~\gev$, and with a lifetime range of about $3\times10^{27} - 2\times10^{28}$~s would be detectable by the Fermi-LAT with a signal-to-noise ratio larger than~3. it has also been shown that gravitino masses larger than about 4 GeV are already excluded in the \mn\ by Fermi-LAT data of the galactic halo~\cite{mnssm-gravitino1}.  
 
$R$-parity breaking couplings can be sufficiently large to lead to interesting expectations for collider phenomenology. The neutralino NLSP, for instance, depending on the RPV model considered, may decay into~\cite{martin,ll-higgsinos}:
\begin{eqnarray} \nonumber
& \qquad   \X\to W^{\pm}\ell^{\mp},       & \qquad \X\to\t{G}\gamma, \\ 
& \qquad   \X\to \nu\tau^{\pm}\ell^{\mp}, & \qquad \X\to\t{G}Z,      \\ \nonumber
& \qquad   \X\to \nu\gamma,               &  \qquad \X\to\t{G}h.
\end{eqnarray}

Such decays may be probed at the LHC via inclusive channels characterised by leptons, many jets, large \met\ and/or photons. The cases where the NLSP is long-lived yet with a decay length comparable to the dimensions of an LHC experiment are particularly interesting, as they give rise to displaced tracks/leptons and non-pointing photons. The possibility to measure the neutralino decay length provides an extra handle to constrain the underlying SUSY model. Nevertheless determining whether $R$~parity is conserved or broken may not be trivial since the DM particles themselves, whether absolutely stable or quasi-stable, cannot be detected directly in collider experiments. To this effect, an interplay between collider and astroparticle physics is necessary to pin down the dark matter properties and the related theoretical-model parameters.

\section{Summary and outlook}\label{sc:summary}

The so-far negative results while searching for supersymmetry in standard channels calls for a more systematic and thorough consideration of non-standard SUSY theoretical models and experimental techniques. To this effect, scenarios involving violations of $R$~parity and/or long-lived particles arise as exciting alternatives.

In particular, bilinear RPV supersymmetry is strongly motivated, as it reproduces correct neutrino physics parameters. In addition to explaining neutrino masses, the $\mu$-from-$\nu$ supersymmetric standard model solves the $\mu$-problem and offers an enriched Higgs and sparticle mass spectrum that easily accommodates a 125-\gev\ Higgs boson. The RPV decays lead to novel signals at colliders: Displaced objects at the LHC, which have less background than other final states and multilepton signatures. More sophisticated searches for displaced objects are expected in near future for ATLAS, CMS and LHCb.

\section*{Acknowledgements}

The author is grateful to the ICNFP2013 organisers for the kind invitation to present this plenary talk in the conference. She acknowledges support by the Spanish Ministry of Economy and Competitiveness (MINECO) under the projects FPA2009-13234-C04-01 and FPA2012-39055-C02-01, by the Generalitat Valenciana through the project PROMETEO~II/2013-017 and by the Spanish National Research Council (CSIC) under the JAE-Doc program co-funded by the European Social Fund (ESF). 


\end{document}